\documentclass [a4paper] {article}

\usepackage[cp1252]{inputenc}
\usepackage[english]{babel}
\usepackage{amssymb}
\usepackage{amsmath}
\usepackage[dvips]{graphicx}

\title{Computational Complexity\\and the Interpretation of a Quantum State Vector}

\author{Arkady Bolotin\footnote{$Email: arkadyv@bgu.ac.il$} \\ \textit{Ben-Gurion University of the Negev, Beersheba (Israel)}}

\begin{document}

\maketitle

\begin{abstract}
\noindent The macro-objectivation problem derives from the fact that the Schrödinger equation is linear and thus requires that a macroscopic system interacting with an entangled state must be entangled as well. However, such a requirement entails that the Schrödinger equation must also be solvable in the macroscopic world in the same way as it is solvable in the microscopic world, which itself is an assumption. In this work, the alternative assumption that the Schrödinger equation is in general an intractable problem is considered.
\end{abstract}

\section{Introduction}

\noindent According to the almost unanimously accepted view, after the famous incompleteness argument by Einstein, Podolsky and Rosen, the question of whether the quantum state vector represents a state of reality or a state of knowledge remains unanswered [1]. Therewith, a connection between this incompleteness argument and the assumption that the Schrödinger equation must be quickly solvable for any given system continues to be unnoticed.\\

\noindent Most tangibly, the link between the incompleteness argument and the assumption of the Schrödinger equation's fast solvability can be demonstrated with the help of the next passage from the paper [2]:

\begin{quotation}
\noindent "Consider a particle counting experiment being conducted by an experimenter while a theoretician does a parallel real time calculation using standard quantum theory. The experimenter sets up the apparatus while the theoretician sets up the initial state vector. While the experimenter turns on the apparatus and monitors its smooth functioning, the theoretician follows the smooth evolution of the state vector according to the Schrödinger equation. Suddenly, the experimenter sings out "An event has occurred, and this is the result.'' Abruptly, the theoretician stops his calculation, replaces the state vector, which by now has become the sum of states corresponding to different possible outcomes of the experiment, by the one state, which the experimenter told him had actually occurred, and then continues his calculation of the smooth evolution of the state vector.\\

\noindent In other words, the practitioner of standard quantum theory must go \textit{outside the theory}, to obtain additional information, in order to use the theory correctly.'' (p. 1-2, italics are in the original)
\end{quotation}

\noindent The conclusion that standard quantum theory does not provide a complete theoretical description is tacitly based here on the assumption that solutions to the Schrödinger equation for a macroscopic system (such as the observing apparatus) must be calculable \textit{quickly}, i.e., within an amount of time that is not greater than the period between the event occurrence and its result receiving. Indeed, if -- despite the rapid rate at which the Schrödinger equation becomes more and more complex as the size of a system increases -- the exact solutions to this equation could be quickly computable even for a truly macroscopic object, then standard quantum theory would be incomplete. Namely, the quantum theory would not be able to describe the process of observation in terms of the state vectors of the observing apparatus and those of the particle under observation since the Schrödinger equation is reversible and deterministic whereas the collapse of the state vector on measurement is irreversible and non-deterministic.\\

\noindent Now, for the sake of argument, assume that while fast calculations of the exact solutions to the Schrödinger equation are possible in very special cases of typically microscopic (atomic) phenomena, in general the Schrödinger equation is an intractable problem, for which there are not any exact solutions reachable in a reasonable amount of time. It would then follow that to reach the exact solution to the Schrödinger equation for the observing apparatus (containing a large number \textit{N} of particles that has the same order of magnitude as Avogadro's number\textit{ N}${}_{A}$) could take the billions or trillions of years regardless of all computing power available to the theoretician.\\

\noindent On the other hand, a problem which requires computer (or mind) crunching would not be so hard if only an approximate answer were needed. Or, putting it differently, the computational cost of the problem solving can be reduced at the cost of statistical uncertainty of the solution. This means that in most cases an intractable problem can be solved quickly only in terms of probability.\\

\noindent Hence, the assumption of the Schrödinger equation's intractability would imply that by the time the experimenter would get the result of the event, the theoretician could only compute the probability of the transition from the state of the particle before the observing interaction to one of the possible states of the particle after the observing interaction corresponding to the different outcomes of the experiment. This would explain why standard quantum theory combines a probabilistic interpretation with deterministic dynamics.\\

\noindent But then again, why would the theoretician be concerned with fast solvability of the Schrödinger equation? Yes, it is certainly true that the enormous complexity of a macroscopic object makes it impossible at present to reach the exact numeric solutions to the Schrödinger equation for the observing apparatus in reasonable time. It is also equally true that exact analytical or closed-form solutions to this equation in the most general case continue to be unknown, even after nearly a century since the equation's first presentation by Dr. Erwin Schrödinger at the Royal Institution, London, in March 1928. And yet, with no evidence disproving principal solvability of the Schrödinger equation for macroscopic systems, it seems quite natural to describe the process of observation quantum-mechanically, i.e., in terms of the state vectors of the observing apparatus and the particle under observation, notwithstanding that at the present time such state vectors can be obtainable only symbolically. For -- even if the theoretician assumes that the general analytical solutions to the Schrödinger equation would never be known -- he could still believe that it is just a matter of time before a powerful numerical method able to quickly compute the exact solutions to this equation for every system is developed.

\section{Computational complexity of the Schrödinger equation}

\noindent The question of fast solvability of the Schrödinger equation is closely related to the question of how fast the resources required to solve this equation grow with the size of the input needed to specify the initial state vector of a system; so before proceeding further, some basic considerations on computational complexity would be in order.\\

\noindent In computational complexity theory, a problem is called \textbf{P} (polynomial) if it can be solved in ``polynomial time'', that is, if an algorithm for solving the problem exists such that the number of steps in the algorithm is bounded by a polynomial function of \textit{n}, where \textit{n} corresponds to the length of the input for the problem. A problem is called \textbf{NP} (nondeterministic polynomial) if the problem's solution can be guessed (and then verified) in polynomial time, but no particular rule is followed how to make the guess.\\

\noindent Examples in the literature of exact (analytical or numerical) solutions to the Schrödinger equation for specific quantum-mechanical systems (see for instance Refs. 3, 4) assure that the Schrödinger equation is a \textbf{NP} problem since all known instances of the equation's correctly guessed solutions have failed to yield the general algorithm for exactly solving this equation in a reasonable number of steps for every system.\\

\noindent Computational complexity theory considers a problem \textbf{NP}-hard if an algorithm for its solution can be modified to solve any \textbf{NP} problem (or any \textbf{P} problem, as \textbf{P} $\subset$ \textbf{NP}). A problem that is both \textbf{NP} and \textbf{NP}-hard is said to be \textbf{NP}-complete. Therefore, finding a fast algorithm for any \textbf{NP}-complete problem implies that a fast algorithm can be found for all \textbf{NP} problems (because a solution for any problem belonging to this class can be modified into a solution for any other member of the class).\\

\noindent Although short in appearance, the Schrödinger equation is immensely powerful. It governs the behavior and properties of all matter at ambient conditions. Therefore, in principle, it can provide a complete description of everything that can happen in the world, including the experimenter and the theoretician as they look at the observing apparatus and learn what the results of the experiment are. This implies that any \textbf{NP} problem that deals with a system obeying the laws of physics can be solved with the Schrödinger equation. In other words, every naturally originated problem in the \textbf{NP} class can be reduced (in a reasonable amount of steps) to the problem of solving the Schrödinger equation associated with the corresponding Hamiltonian.\\

\noindent It is easy to demonstrate that among all naturally originated \textbf{NP} problems there exists at least one that is \textbf{NP}-hard. As an illustration, the wave function of a system with large number of strongly interacting fermions changes sign when any two fermions are interchanged (due to the symmetry of the wave function). For this reason, the sum over all multi-particle states involves an integral over a function that is highly oscillatory and hence hard to calculate (particularly in high dimension, i.e., in the thermodynamic limit since the dimension of the integral is given by the number of particles). Consequently, to describe any fermionic system in terms of the exact solutions to the Schrödinger equation connotes to find a full and generic solution of the ``sign problem'' that has been proven \textbf{NP}-hard [5].\\

\noindent As an inference from this fact, it follows that the Schrödinger equation belongs to the \textbf{NP}-complete class. This means that finding a fast algorithm for the exact solutions of the Schrödinger equation in the most general case would imply that a fast exact generic algorithm could be found for all \textbf{NP} problems including those that deal with pure mathematical operations and logical constructions.\\

\noindent Since the universal analytical or numerical algorithm specifying how to exactly solve the Schrödinger equation for each system is unknown, in order to solve the Schrödinger equation for a particular system one is forced to use brute-force methods (so called exhaustive search). However, given that the dimension of configuration space grows linearly with the number \textit{N} of particles constituting the system, the computational effort to solve the Schrödinger equation would scale exponentially with the constituent particle number \textit{N}. The \textbf{P} versus \textbf{NP} question (the single most important question in all of theoretical computer science and one of the most important in all of mathematics [7]) asks then whether this exhaustive search over an exponentially large set can be avoided in general, i.e., for every system.\\

\subsection{What if the Schrödinger equation is solvable fast?}

\noindent If the assumption of fast solvability of the Schrödinger equation is correct, then \textbf{P} = \textbf{NP}. That is, the generic algorithm that is able to quickly compute the exact solutions to the Schrödinger equation for any system does exist, and hence it will be certainly found someday. And when it happens, the world will be like a Utopia. Since every natural science will be derivable in reasonable amount of steps from quantum mechanics, the view that the nature is arranged into the layers of increasing complexity with each requiring its own special science will be irrelevant. For example, fast solutions to any classical mechanical problems will be obtained from quantum mechanics in the approximation that the Planck's constant $\hbar $ is zero, and the properties of all molecules regardless of their complexity will be quickly computed from Schrödinger equation. Every explanation in biology and psychology will be derivable in reasonable amount of steps from the exact and quickly reachable solutions to the many-body Schrödinger equation. Every macroscopic object -- the observing apparatus, the earth, the whole universe -- will be treated in terms verifiable practically (i.e., not just symbolically) as a quantum system. What's more, due to \textbf{NP}-completeness of the Schrödinger equation, its fast solutions will be modified (in reasonable amount of steps) to quickly solve all pure mathematical and logical \textbf{NP} problems. This will lead, in Gödel's words (see Ref. 6), to ``consequences of the greatest magnitude'' completely replacing mental effort of mathematicians by machines.\\

\noindent However, this Utopian world will come at a price. Incompleteness of standard quantum theory will be a proven fact inasmuch as the standard interpretation cannot jointly incorporate two conditions: first, that a measurement requires an interaction with an `outside' (macroscopic) system; and, secondly, that the whole universe can be treated as a quantum system [8].\\

\subsection{What if the Schrödinger equation is intractable?}

\noindent If in most cases the Schrödinger equation is intractable, then no \textbf{NP}-complete problem has a fast algorithm (otherwise, such an algorithm could be used to quickly solve the Schrödinger equation in every case), and thus \textbf{P} $\neq$ \textbf{NP}.\\

\noindent In these circumstances, the existence of different theories organized into a family tree makes sense: Though principally every theory is derivable from the Schrödinger equation, in practice deriving, say, chemistry or biology from this equation would be absolutely hopeless. Thus, the further down the tree from quantum mechanics, the more empirical (or approximate) approach is taken by a theory.\\

\noindent Because to reach the exact solution to the Schrödinger equation will in most cases take an exponential amount of steps, to do an exact calculation using the Schrödinger equation \textit{at the same speed as an experiment} will be an unfeasible task except for small inputs (such as microscopic systems totally isolated from the environment) or cases with some special properties (such as the collective behavior of constituent particles that decreases dramatically the complexity of the many-body Schrödinger equation). This implies that the von Neumann chain, which has its seed at the microscopic level, will not be able to develop into any practically computable (and hence verifiable) macroscopic consequence (like the Schrödinger cat in an impossible state) since when the chain of entangled systems reaches the macroscopic level (with its enormous number of degrees of freedom) the Schrödinger equation will become unsolvable for all practical purposes. Hence, the impossibility of deriving a nonsymbolical description of a macroscopic system from the Schrödinger equation is the reason for the world (or just its portrayal) splitting into a macroscopic domain following classical mechanics and a microscopic domain following quantum mechanics.\\

\noindent Due to its asymptotic character, the assumption of the Schrödinger equation's intractability does not suggest any precise criterion (based on physical properties of the systems involved) for identifying the borderline between quantum and classical descriptions. This might explain why the split between micro and macro is fundamentally shifty.\\

\noindent A problem that requires taking exponential time to find the exact solution might allow for a fast approximation algorithm that returns a solution with a degree of statistical uncertainty. In fact, allowing an algorithm to make mistakes (i.e., allowing statistical uncertainty in the solution) is equivalent to making the description of a system \textit{coarser} (less detailed), which consequently might allow the algorithm to run in polynomial time rather than exponential time.\\

\noindent This means that instead of the exact solution to the Schrödinger equation precisely describing every particle of the observing apparatus (but not reachable before the Sun goes cold), the theoretician might quickly (in real time, that is, during the duration of the experiment) find a probabilistic solution ignoring the detector's microscopic degrees of freedom.\\

\noindent For example, assume that the initial state vector is the superposition of two states that are only possible for the observed particle:

\begin{equation} \label{1} 
\left.\left|\psi \left(0\right)\right.\right\rangle =\frac{1}{\sqrt{2}}\left.\left|1\right.\right\rangle +\ \frac{1}{\sqrt{2}}\left.\left|2\right.\right\rangle   .  
\end{equation}

\noindent If the observing interaction happens at the moment $t=0$, then in the next moment $\delta t$ the state vector will be specified by the unitary evolution

\begin{equation} \label{2}
\left.\left|\psi \left(\delta t\right)\right.\right\rangle =U\left.\left|\psi \left(0\right)\right.\right\rangle =\left(I-\frac{i}{\hbar}H\delta t\right)\left.\left|\psi \left(0\right)\right.\right\rangle
\end{equation}

\noindent defying (mostly) by the interaction Hamiltonian $H\approx H_{{\rm int}}$

\begin{equation} \label{3}
H_{{\rm int}}=\left(\left.\left|1\right.\right\rangle\left\langle\left.1\right|\right.\right)\otimes \left(\sum_k{A_{1k}\left.\left|\epsilon_k\right.\right\rangle \left\langle\left.\epsilon_k\right|\right.}\right)+\left(\left.\left|2\right.\right\rangle\left\langle\left.2\right|\right.\right)\otimes \left(\sum_k{A_{2k}\left.\left|\epsilon_k\right.\right\rangle \left\langle \left.\epsilon_k\right|\right.}\right)   ,
\end{equation}

\noindent where $\left.\left|\epsilon_k\right.\right\rangle$ are the orthonormal basis vectors for the observing apparatus (i.e., the exact solutions to the Schrödinger equation describing each configuration of the apparatus's constituent particles), $A_{1k}$ and $A_{2k}$ are the corresponding interaction coefficients. Since there is no hope to find the exact solutions $\left.\left|\epsilon_k\right.\right\rangle $ in real time, to follow the experiment the theoretician might choose the coarse graining description of the observing apparatus that ignores the individual configurations of the apparatus's microscopic particles:

\begin{equation} \label{4}
H_{{\rm int}}\cong \left(\left.\left|1\right.\right\rangle \left\langle \left.1\right|\right.\right) \left({\overline{A}}_1\pm \triangle A_1\right)+\left(\left.\left|2\right.\right\rangle \left\langle \left.2\right|\right.\right) \left({\overline{A}}_2\pm \triangle A_2\right)   ,
\end{equation}

\noindent where ${\overline{A}}_1$ and ${\overline{A}}_2$ are the estimated interaction coefficients, $\triangle A_1$ and $\triangle A_2$ are their uncertainties. A stochastic unitary process defined by the Hamiltonian (4) will give the set of possible (nondeterministic) solutions at the moment $\delta t$

\begin{equation} \label{5}
\left.\left|\psi \left(\delta t\right)\right.\right\rangle \in \left\{\frac{\left.\left|1\right.\right\rangle }{\sqrt{2}}{ \exp\left[-\frac{i\left({\overline{A}}_1\pm \triangle A_1\right)\delta t}{\hbar }\right]\ }+\frac{\left.\left|2\right.\right\rangle }{\sqrt{2}}{\exp \left[-\frac{i\left({\overline{A}}_2\pm\triangle A_2\right)\delta t}{\hbar }\right]\ \ }\right\} .
\end{equation}

\noindent The probability of the transition from the initial state (1) before the observing interaction to one of the possible states (5) after the observing interaction will be

\begin{displaymath}
P\left(0\to \delta t\right)={\left|\left\langle \psi \left(0\right)\mathrel{\left|\vphantom{\psi \left(0\right) \psi \left(\delta t\right)}\right.\kern-\nulldelimiterspace}\psi \left(\delta t\right)\right\rangle \right|}^2\in 
\end{displaymath}
\begin{equation} \label{6}
\in\left\{\frac{1}{4}{\left|{\exp  \left[-\frac{i\left({\overline{A}}_1\pm \triangle A_1\right)\delta t}{\hbar }\right]\ }+{\exp  \left[-\frac{i\left({\overline{A}}_2\pm \triangle A_2\right)\delta t}{\hbar }\right]\ \ }\right|}^2\right\}   .
\end{equation}

\noindent To calculate this probability, the theoretician has to first generate inputs $\triangle A_1$ and $\triangle A_2$ randomly from probability distributions over the domains of possible values for $\triangle A_1$ and $\triangle A_2$, then perform a deterministic computation with (6) on these inputs, and finally aggregate the results. In this way the theoretician will get

\begin{displaymath} 
P\left(0\to \delta t\right)=
\end{displaymath}
\begin{displaymath}
=\frac{1}{2}  + \frac{1}{2} \left({\cos\frac{ { \overline{A} }_1 - { \overline{A} }_2} {\hbar }\delta t\ } {\overline{{\cos  \frac{\pm \triangle A_1\mp \triangle A_2}{\hbar }\delta t\ }}} \right) - 
\end{displaymath}
\begin{equation} \label{7}
-\frac{1}{2} \left({\sin \frac{ {\overline{A}}_1 - {\overline{A}}_2} {\hbar }\delta t\ } {\rm \ } {\overline{{\sin  \frac{\pm \triangle A_1\mp \triangle A_2}{\hbar }\delta t\ }}} \right)=\frac{1}{2} 
\end{equation}

\noindent since the average value of the cosine and the sine taking on several random inputs $\triangle A_1$ and $\triangle A_2$ will be zero. This aggregate result means that after the observing interaction, the initial sum of states (1) corresponding to two possible outcomes of the experiment will collapse and the observed particle will be found with equal probability in one of the possible states.\\

\section{Discussion}

\noindent As the studies on the foundations of quantum mechanics reveal, the largely debated macro-objectivation problem (concerning the transition from the microscopic quantum world to the macroscopic world perfectly described by classical mechanics) is actually based on one tacit assumption common to a great degree for theoretical physics on the whole.\\

\noindent Indeed, the macro-objectivation problem derives from the fact that the Schrödin-ger equation is linear and thus requires that a macroscopic system interacting with an entangled state must be entangled as well. However, such a requirement entails that the Schrödinger equation must also be solvable in the macroscopic world in the same way as it is solvable in the microscopic world, which itself is an assumption. In Alan Turing's words, this is the assumption ``that as soon as a fact is presented to a mind all consequences of that fact spring into the mind simultaneously with it. It is a very useful assumption under many circumstances, but one too easily forgets that it is false'' [9]. That is, this is the assumption that as soon as the theoretician writes down a symbolic form of the Schrödinger equation providing the description of the experiment, he immediately gets all the consequences and predictions. This means that the belief in solvability of the Schrödinger equation in the macroscopic world either does not take into account the resources required to solve this equation (considering such resources unlimited) or presumes that this equation is fast solvable for each system. But both such justifications can be wrong -- the first is because it is unrealistic, the second is because \textbf{P} might be not equal to \textbf{NP}.\\

\noindent What surprising is here is that at the same time as most computer scientists are trying to prove that \textbf{P} $\neq$ \textbf{NP}, the majority of the physicists engaging in the studies on the foundations of quantum mechanics do not question the solvability of the Schrödinger equation in the macroscopic world and so in effect believe that \textbf{P} = \textbf{NP}.\\

\section*{References}

\begin{enumerate}
\item  G. Ghirardi and R. Romano\textit{, On the Completeness of Quantum Mechanics and the Interpretation of the State Vector}, arXiv:1302.6278v1 [quant-ph] 25 Feb 2013.

\item  P. Pearle, \textit{True Collapse and False Collapse}, Published in Quantum Classical Correspondence: Proceedings of the 4th Drexel Symposium on Quantum Non-integrability, Philadelphia, PA, USA, September 8-11, 1994, pp. 51-68.

\item  A. Alhaidari, \textit{Exact solutions of Dirac and Schrödinger equations for a large class of power-law potentials at zero energy}, Int. J. Mod. Phys. A, 17, 45-51, arXiv: 0112001 (2002).

\item  R. Zhang, C. Deng, \textit{Exact solutions of the Schrodinger equation for some quantum-mechanical many-body systems}, Phys. Rev. A, 47, 1, 71-77 (1993).

\item  M Troyer, U.-J. Wiese, \textit{Computational complexity and fundamental limitations to fermionic quantum Monte Carlo simulations}, arXiv: 0408370v1 (2004).

\item  M. Sipser. \textit{The history and status of the P versus NP question}, Proc. ACM STOC, 603--618 (1992).

\item  L. Fortnow, \textit{The Status of the P versus NP Problem}, Communications of the ACM, 52, 9, 78-86 (2009).

\item  H. Putnam, \textit{Comments on the paper of David Sharp}, Philosophy of Science, 28, 234--237 (1961).

\item  A. Turing. \textit{Computing machinery and intelligence}. Mind, 59:433--460, 1950.
\end{enumerate}

\end{document}